\documentclass[twocolumn,pra]{revtex4}%
\usepackage{amsmath}
\usepackage{amssymb}
\usepackage{graphicx}
\usepackage{url}
\usepackage{amsfonts}%
\DeclareMathAlphabet{\mathwee}{OT1}{cmss}{m}{sl}

\begin{document}
\title{A hybrid on-chip opto-nanomechanical transducer for ultra-sensitive force measurements}
\author{E.\ Gavartin$^{1}$, P.\ Verlot$^{1}$, T.\ J.\ Kippenberg$^{1,2}$}
\affiliation{$^{1}$ Ecole Polytechnique F{\'e}d{\'e}rale de Lausanne,
EPFL, 1015 Lausanne, Switzerland\\
$^{2}$ Max-Planck-Institut f{\"u}r Quantenoptik,
Hans-Kopfermann-Stra{\ss}e 1, 85748 Garching, Germany}

\small

\begin{abstract}
\bf{
\noindent
Nanomechanical oscillators have been employed as transducers to
measure force, mass and charge with high sensitivity. They are also used in
opto- or electromechanical experiments with the goal of quantum control and phenomena of mechanical systems. Here, we report the realization and
operation of a hybrid monolithically integrated transducer system consisting of a
high-$Q$ nanomechanical oscillator with modes in the MHz regime coupled to the
near-field of a high-$Q$ optical whispering-gallery-mode microresonator. The transducer system enables a sensitive resolution of the nanomechanical beam's thermal motion with a signal-to-noise of five orders of magnitude and has a
force sensitivity of $74\,\rm{aN}\,\rm{Hz}^{-1/2}$ at room temperature. We show, both theoretically and
experimentally, that the sensitivity of continuous incoherent
force detection improves only with the fourth root of the averaging time. Using dissipative feedback based on radiation pressure enabled control, we explicitly demonstrate by detecting a weak incoherent force that this constraint can be significantly relaxed. We achieve a more than 30-fold reduction in averaging time with our hybrid transducer and are able to detect an incoherent force having a force spectral density as small as $15\,\rm{aN}\,\rm{Hz}^{-1/2}$ within $35\,\rm{s}$ of
averaging. This corresponds to a signal which is 25 times smaller than the thermal noise and would otherwise remain out of reach. The reported monolithic platform is an enabling step towards hybrid nanomechanical
transducers relying on the light-mechanics interface.}

\end{abstract}

\maketitle

Nanomechanical oscillators \cite{Ekinci2005} serve as ultrasensitive detectors
of force \cite{Mamin2001}, mass \cite{Jensen2008} and charge
\cite{Cleland1998}. Recently, increasing efforts have
been devoted to sensitively detect the nanomechanical motion of these oscillators
\cite{LaHaye2004, Knobel2003, Poggio2008, Etaki2008} with recent systems exhibiting a sensitivity
below that at the standard quantum limit (SQL) \cite{Teufel2009,Anetsberger2010}.  For sensitive force detection the requirements on displacement sensitivity are less stringent, though, because of the thermal limit. Recent work has demonstrated that trapped ions have also the potential to be employed as sensitive transducers for ultrasmall forces \cite{Biercuk2010, Knuenz2010}, with the force sensitivity reaching levels of only $5\,\rm{yN}\,\rm{Hz}^{-1/2}$ \cite{Knuenz2010}. While being a promising approach to detect specific small forces, it presently suffers from challenging technology required for ion trap experiments, stable interaction times being only in the millisecond range \cite{Biercuk2010} and a lack of field-deployable sensors. In contrast, cantilever-based sensing is a well-established technique that allowed remarkable achievements such as the detection of single spins \cite{Rugar2004} and the reconstruction of the structure of a virus \cite{Degen2009}, and it is increasingly used in the emerging field of biosensing \cite{Roukes2011}. A particularly promising approach is to parametrically couple a nanomechanical oscillator to a microwave \cite{Teufel2009,Regal2008} or optical
\cite{Anetsberger2009} microcavity, thus enhancing the displacement sensitivity and allowing to efficiently resolve the motion of cantilevers with dimensions below the diffraction limit. Previous realizations suffered, however, from the necessity of cryogenic operation \cite{Teufel2009,Regal2008}, from the fact that the optical resonator
and mechanical oscillator were realized on two different chips and had to be
brought to close proximity by external positioners \cite{Anetsberger2009} making the system too unstable for usage as a force transducer or from low mechanical $Q_{\rm{{M}}}$, significantly limiting the force resolution \cite{Kartik2011, Li2009}. Here, we present an integrated hybrid transducer system consisting of a high-stress Si$_{3}$N$_{4}$ nanomechanical beam whose high-$Q_{\rm{M}}$  out-of-plane mode is coupled to the near-field of a whispering gallery mode of a silica disk resonator. As such, we combine the arguably best material for high-$Q_{\rm{M}}$ nanomechanics \cite{Verbridge2006} (stoichiometric Si$_{3}$N$_{4}$) with an advantageous material for linear optics \cite{Braginsky1989,Vernooy1998,Armani2003,Kippenberg2006} (SiO$_{2}$) in an integrated on-chip device, which allows the detection of ultra-small forces at the aN level.  The advantages of silica for optical resonators result from higher thresholds for thermal nonlinearities when compared to silicon \cite{Borselli2005} or silicon nitride \cite{Gondarenko2009} as well as the absence of two photon absorption \cite{Johnson2006}, thus allowing significantly higher circulating power and preventing cross talk between optical modes induced by the nonlinearity.  Moreover, silica microresonators can be coupled with high efficiency to optical fiber as required for highly efficient measurements of position or force.

The versatility of the presented hybrid transducer is enhanced by implementing a feedback scheme that introduces an
effective susceptibility for the mechanical resonator
\cite{Mertz1993,Mancini1998,Cohadon1999}. This enables tuning of the mechanical resonance
frequency and linewidth. This feature is particularly interesting in the
context of ultra-sensitive force detection, which relies on the use of
high-$Q_{\rm{M}}$ mechanical oscillators whose detection bandwidth is limited to
a very narrow range around their resonance frequency
\cite{Mertz1993,oosterkamp2010}. Here,
we report the implementation of an integrated feedback scheme enabling full and
independent control of the effective mechanical properties of the transducer. Though being
limited mostly by thermal noise, nanomechanical transducers
can be used to detect arbitrarily weak coherent signals, with the sensitivity scaling with the square root of the averaging
time $\tau$ \cite{Rugar2004}. The situation is more delicate for detecting
incoherent signals, such as quantum backaction \cite{verlot2009scheme},
atomic mass shot noise \cite{Jensen2008}, or a gravitational wave background
\cite{Abbott2009}. Here the absence of phase coherence for such signals
prevents using amplitude averaging to retrieve their signature, therefore energy
averaging must be employed \cite{Rugar2004}. However, this is made at the
expense of the measurement resolution, which increases only very slowly
with $\tau^{1/4}$ \cite{Rugar2004}. Here, we propose and experimentally demonstrate with our hybrid transducer that dissipative feedback allows a significant decrease of the averaging time required to
reach a given accuracy in the context of a low signal-to-noise ratio (SNR)
force measurement.\\

\noindent \textbf{Near-field coupled hybrid transducer.}
Our integrated system enables hybrid integration of the approach of Anetsberger \textit{et al.}
\cite{Anetsberger2009} and is fabricated using a simple procedure easy to be implemented for mass production (fabrication details are given in Methods). Images of the structure are given in Figures
1(a) and 1(b). It consists of a
doubly-clamped, high-stress Si$_{3}$N$_{4}$ nanomechanical beam
\cite{Verbridge2006} (blue) which is realized above the wedge of a disk made
of SiO$_{2}$ (red). Both, the pads to which the oscillator is clamped and the
silica disk rest on silicon pedestals (green). We choose this geometry to optimize the hybrid transducer's performance as a force sensor, which is directly related to the mechanical quality factor $Q_{\rm{M}}$ and mass of the cantilever. It was shown \cite{Verbridge2006} that the fundamental out-of-plane mode of a Si$_{3}$N$_{4}$ nano-beam exhibits distinctly high $Q_{\rm{M}}$ factors while having a small mass, thus making it an optimal candidate for a force transducer. In order to efficiently couple the out-of-plane motion of the beam to the evanescent field of a WGM disk cavity, the beam needs to be positioned above the wedge of the disk resonator, where the distribution of the near-field electromagnetic energy is maximized (cf. Fig. 1(d)). The beam of the hybrid system used in this work has dimensions of
$90\;\mu\mathrm{m}\times700\;\mathrm{nm}\times100\;\mathrm{nm}$ and the disk
has a diameter of 76$\;\mu\mathrm{m}$, with the distance between the beam and
the disk being approximately $250$ nm. Like microtoroidal cavities \cite{Armani2003} silica microdisk
cavities support high-$Q$ optical whispering gallery modes (WGMs)
\cite{Kippenberg2003}. In our system we measured different families of WGMs
around a wavelength of 1550 nm by coupling laser light coming from an external
cavity diode laser to the cavity via the fiber-taper technique \cite{Cai2000} (Fig. 1 (c)).
The linewidth of such modes is around $\delta\nu\approx3\;\mathrm{GHz}$ corresponding to a value of $Q=\nu/\delta\nu_{0}\approx65,000$ with
$\nu_{0}$ being the resonance frequency of the mode. This measured linewidth is
considerably higher when compared to previous results
\cite{Kippenberg2006,Kippenberg2003}, which is explained by enhanced scattering
losses due to the defects present on the disc arising from shadowing effects during fabrication. While the increased optical linewidth limits the attained \textit{displacement} sensitivity of the mechanical motion to the level of several tens of $\rm{fm}\,\rm{Hz}^{-1/2}$, the \textit{force} sensitivity is not affected, since the thermally driven motion of the mechanical transducer is well resolved as shown below. In fact, an improved linewidth would even be detrimental for the system's ability to function as a force transducer due to effects of \textit{dynamical backaction} \cite{Kippenberg2008}, which would affect the stability of the mechanical transducer.

To measure the transduction of the mechanical motion of the oscillator, we keep the hybrid system under a pressure of below $10^{-4}$ mbar and detune the laser to the blue side of the fringe of a critically coupled optical mode
\cite{Cai2000}. We are able to resolve thermally driven mechanical motion of the nanomechanical oscillator, which is imprinted as a modulation on the transmitted optical intensity. Launching a laser
power of 400 $\mu\mathrm{W}$ into the cavity, we observe the fundamental
out-of-plane mode of the beam at a resonance frequency of $\Omega_{\rm{M}}%
/2\pi=2.88\;\mathrm{MHz}$ with a signal-to-noise ratio of 50 dB. By applying a known phase modulation \cite{Gorodetsky2010} we can determine the optomechanical coupling of the system to be $G/2\pi=
d\nu /dx=2.9\;\mathrm{MHz/nm}$ ($x$ is the displacement of the mechanical oscillator) and calibrate the spectrum in frequency
noise $S_{\nu\nu}\left[  \Omega\right] $ as well as displacement density $S_{xx}\left[
\Omega\right]  =S_{\nu\nu}\left[  \Omega\right]  /G^{2}$ units. The effective mass \cite{Pinard1999} $m_{\mathrm{eff}}$ of the
fundamental out-of-plane mechanical mode coupled to a WGM optical mode equals half the physical mass $m$ \cite{Anetsberger2009}, thus giving $m_{\mathrm{eff}}=9\times10^{-15}\;\mathrm{kg}$ for our system. By fitting the data we
obtain a very high mechanical quality factor of $Q_{\mathrm{{M}}}%
=4.3\times10^{5}$. To confirm this value, we perform a ringdown measurement
of the oscillator by actuating the beam at $\Omega_{\rm{M}}$ via the radiation
pressure force and interrupting the actuation while recording the response (scheme is shown in Fig. 2(a)). We obtain a value of
$4.8\times10^{5}$ (Fig. 2(b)) corroborating the spectrally obtained result. To exclude dynamical backaction effects via radiation pressure in our measurements, we varied laser power and detuning around their working points and did not observe any change in $\Omega_{\rm{M}}$ or $Q_{\rm{M}}$, ensuring that this type of interaction is indeed weak.\\

\noindent \textbf{Actuation via the radiation pressure force.}
To demonstrate the actuation of the mechanical oscillator via the radiation
pressure force, we perform a force response measurement using a pump-probe
scheme. The setup is shown in Fig. 2(a) and details of the experiment are
given in Methods. Fig. 2(c) shows the response of the mechanical oscillator to different
levels of actuation. The response increases with stronger actuation and shows
a nonlinear behavior above a given threshold that can be described by the
Duffing model \cite{Nayfeh1979}.\\

\noindent \textbf{Feedback tuning and cooling of the transducer.}
We implement a feedback scheme \cite{Mancini1998} to effectively address the mechanical
properties of our nanomechanical oscillator, which, as shown later, is important for an improved force detection. Feedback consists in applying
a force $F_{\mathrm{{fb}}}\left[\Omega\right]$ to the oscillator proportional to its displacement $x$. In
Fourier space this force can be written as:%

\begin{align}
F_{\mathrm{{fb}}}[\Omega]=-g e^{i \Phi}\times m_{\mathrm{{eff}}}%
\Gamma_{\mathrm{{M}}}\Omega_{\mathrm{{M}}}x[\Omega],\label{eq:force-fb}%
\end{align}
where $g$ and $\Phi$ denote the amplitude and the phase of the gain,
respectively. Thus, in presence of feedback, the response of the oscillator to
an external force $F_{\mathrm{{ext}}}$ reads:%

\begin{align}
x[\Omega] & =\chi[\Omega]\left( F_{\mathrm{{ext}}}\left[\Omega\right]+F_{\mathrm{{fb}}}\left[\Omega\right]\right)
\nonumber\\
& =\chi_{\mathrm{{eff}}}[\Omega]F_{\mathrm{{ext}}}[\Omega
],\label{eq:response-fb}%
\end{align}
where $\chi_{\mathrm{{eff}}}$ denotes the (Lorentzian) effective
susceptibility \cite{arcizet2006high}, characterized by effective mechanical
resonance frequency $\Omega_{\mathrm{{eff}}}$ and effective damping
$\Gamma_{\mathrm{{eff}}}$ given by:
\begin{align}
\Omega_{\mathrm{{eff}}}^{2} & =\Omega_{\mathrm{{M}}}^{2}\left( 1+\frac
{g}{Q_{\mathrm{{M}}}}\cos{\Phi}\right) \label{eq:effective-omega}\\
\Gamma_{\mathrm{{eff}}} & =\Gamma_{\mathrm{{M}}}\left( 1+g\sin{\Phi}\right)
.\label{eq:effective-gamma}%
\end{align}
The real part of the gain $g \cos{\Phi}$ determines the conservative
contribution of the feedback force modifying the effective mechanical
resonance frequency of the oscillator as shown by Equation \ref{eq:effective-omega}. Equivalently, the imaginary part of the gain $g\sin{\Phi}$ sets the
dissipative component of the feedback force changing the effective mechanical
damping rate of the oscillator (Eq. \ref{eq:effective-gamma}). Consequently,
independent control of amplitude $g$ and phase $\Phi$ of the feedback enables
tuning the oscillator's mechanical response.

Historically feedback implementation has been based on the use of resonant
electrical circuits designed to create a feedback signal at the mechanical
frequency of interest \cite{Cohadon1999,Bouwmeester2006,Poggio2007}. This
method suffers from the difficulty to tune these circuits, so that the actual
feedback implementation is limited to one specific mechanical mode. Here, we use a method based on a demodulation-modulation technique \cite{Poot2011}, which enables accurate control of any mechanical mode
up to arbitrary high frequencies.  The scheme employed is shown in Figure 3(b). The method relies on reading out the time evolution of the
quadratures $X_{1}$ and $X_{2}$, which are the slowly
varying components of mechanical motion $x(t)$ around its resonance frequency
$\Omega_{\mathrm{{M}}}$,%

\begin{align}
x(t)\simeq X_{1}(t)\cos{\Omega_{\mathrm{{M}}}t}+X_{2}(t)\sin{\Omega_{\mathrm{{M}}}t}.\label{eq:quadratures}%
\end{align}

The error signal is obtained by mixing these quadratures at the mechanical resonance frequency $\Omega_{\rm{M}}$ with tunable phase and amplitude (see Methods), and is fed back into the cavity via amplitude modulation the pump beam.

Using this scheme we demonstrate accurate feedback control of the
fundamental out-of-plane mode by tuning $\Phi$. Figure 3(c) shows the feedback
induced resonance frequency shift $(\Omega_{\mathrm{{eff}}}-\Omega
_{\mathrm{{M}}})/2\pi$ (left axis, blue dots) and normalized effective damping
$\Gamma_{\mathrm{{eff}}}/\Gamma_{\mathrm{{M}}}$ (right axis, red dots) versus
the feedback phase $\Phi$ with a constant $g$. The experimental data of Figure
3(c) is fitted using Equations. \ref{eq:effective-omega} and
\ref{eq:effective-gamma} giving excellent agreement. In both cases we use the
same fitting parameter $g=30.7$, thus demonstrating the reliability of our scheme.

We particularly investigate the case $\Phi=\pi/2$, which corresponds to
purely dissipative feedback and has recently received increased interest as it
enables to cool a mechanical degree of freedom
\cite{Cohadon1999,Courty2001,Bouwmeester2006,Poggio2007,Raizen2011}. We
apply dissipative feedback with different gain amplitudes to the fundamental
out-of-plane mode of the oscillator (cf. Fig. 3(a)). In
presence of feedback the oscillator is described by thermodynamical
equilibrium as long as the spectral density of thermal motion $S_{\mathrm{{xx}%
}}^{\mathrm{{fb}}}[\Omega\simeq\Omega_{\mathrm{{M}}}]$ is larger than the
readout imprecision $S_{\mathrm{{r}}}[\Omega\simeq\Omega_{\mathrm{{M}}}]$. In this limit an effective temperature
$T_{\mathrm{{eff}}}$ can be defined as $T_{\mathrm{{eff}}}%
=(\Gamma_{\mathrm{{M}}}/\Gamma_{\mathrm{{eff}}})T$. Thus, starting from room
temperature, we achieve a minimum effective temperature $T_{\mathrm{eff}%
}=700\;\mathrm{mK}$, limited only by the readout sensitivity.  To demonstrate the versatility
of our method, we perform feedback cooling on the third
order harmonic of the out-of-plane mode by simply changing the
modulation/demodulation frequency. The results are shown in Fig. 3(d)
and demonstrate that the scheme can be efficiently used for different modes.\\

\noindent \textbf{Force resolution enhancement via feedback control.}
High-$Q$ nanomechanical oscillators, combined with high sensitivity readout, are ideally suited for measurements of weak forces \cite{Mamin2001,Teufel2009}. Their sensitivity is mainly limited by the thermal Langevin force $F_{\rm{th}}$ with the spectral density (SD) given by the fluctuation-dissipation theorem $S_{\rm{FF}}^{\rm{th}}[\Omega]=4 k_{\rm{B}}T/\Omega\,\rm{Im}\left(1/\chi[\Omega]\right)$, where $k_{\rm{B}}$ is Boltzmann's constant. In the limit of high-$Q_{\rm{M}}$ harmonic oscillators, the SD reads $S_{\rm{FF}}^{\rm{th}}[\Omega]\simeq4 m_{\rm{eff}} k_{\rm{B}}T\Gamma_{\rm{M}}$ and is proportional to both the mass and mechanical damping. For our integrated system this sensitivity is $S_{\rm{FF}}^{\rm{th}}[\Omega]\simeq (74\,\rm{aN})^2/\rm{Hz}$, comparable to the best values reported at room temperature for cantilever-based systems \cite{yasumura2000quality}, however with the benefit of an almost two orders of magnitude higher bandwidth \cite{Mertz1993}. In principle, detecting stationary incoherent forces embedded in the thermal noise is possible by means of energy averaging. However, this averaging converges very inefficiently, scaling with $\tau^{1/4}$ \cite{Rugar2004} and thereby limiting the force detection threshold. In the following, we present a detailed study of energy averaging, and show that its convergence can be drastically improved by using dissipative feedback.

We are interested in detecting a signal $x_{\rm{sig}}(t)$ driven by a stationary incoherent force $\delta F_{\rm{sig}}(t)$, which is uncorrelated with the thermal force and for which $S_{\rm{FF}}^{\rm{sig}}[\Omega\simeq\Omega_{\rm{m}}]\ll S_{\rm{FF}}^{\rm{th}}[\Omega\simeq\Omega_{\rm{m}}]$ holds true, i.e. the incoherent force is buried in the thermal noise. In presence of $\delta F_{\rm{sig}}$, the energy of the transducer $\mathcal{E}_{\mathrm{m}}=\frac{1}{2}m_{\mathrm{eff}}\Omega_{\rm{m}}^2\langle x^2\rangle$ can be written as the contribution of two independent terms $\mathcal{E}_{\mathrm{th}}=\frac{1}{2}m_{\mathrm{eff}}\Omega_{\rm{m}}^2\langle x_{\rm{th}}^2\rangle$ and $\mathcal{E}_{\mathrm{sig}}=\frac{1}{2}m_{\mathrm{eff}}\Omega_{\rm{m}}^2\langle x_{\rm{sig}}^2\rangle$, where $\langle...\rangle$ denotes the average over the statistical domain describing $x$ as a random variable. Thus, comparing the transducer's energy with and without $\delta F_{\rm{sig}}$ being applied enables in principle its detection. Experimentally, statistical averages are not available, and one has to rely on estimators. Assuming $\delta F_{\rm{sig}}$ to be a Gaussian white noise for simplicity, $s(\tau)=1/\tau\int_0^\tau dt X^2(t)$ is a non-biased estimator for the measurement of the energy, with $X$ denoting any of the motion's quadratures. $X(t)$ is the sum of two uncorrelated contributions $X_{\rm{th}}(t)$ and $X_{\rm{sig}}(t)$, giving therefore 

\begin{eqnarray}
	s(\tau)=s_{\rm{th-th}}(\tau)+2s_{\rm{th-sig}}(\tau)+s_{\rm{sig-sig}}(\tau)
\end{eqnarray}

with $s_{\mathrm{i-j}\in\{\rm{th,sig}\}}(\tau)=1/\tau\int_0^\tau dt X_{\rm{i}}(t)X_{\rm{j}}(t)$. If we assume the measurement sampling time $\tau_{\rm{s}}$ to be small compared to the mechanical coherence time $\tau_{\rm{s}}\ll 1/\Gamma_{\rm{M}}$, the statistical dispersion of each of these $3$ terms can be calculated \cite{verlot2011towards} to be:

\begin{eqnarray}
	\Delta s_{\rm{i-i}}(\tau)&=&\sqrt{2}H(\Gamma_{\rm{M}},\tau)\langle X_{\rm{i}}^2\rangle,\label{eq:dispersions1}\\
    \Delta s_{\rm{th-sig}}(\tau)&=&H(\Gamma_{\rm{M}},\tau)\langle X_{\rm{th}}^2\rangle^{\frac{1}{2}} \langle X_{\rm{sig}}^2\rangle^{\frac{1}{2}},
	\label{eq:dispersions2}
\end{eqnarray}

where $H(\Gamma_{\rm{M}},\tau)=\sqrt{2}\left(e^{-\Gamma_{\rm{M}}\tau}-1+\Gamma_{\rm{M}}\tau\right)^{\frac{1}{2}}/\left(\Gamma_{\rm{M}}\tau\right)$ depends only on the mechanical linewidth of the transducer. Noting that $\langle X_{\rm{i}}^2\rangle\simeq S_{\rm{FF}}^{\mathrm{i}}[\Omega_{\rm{M}}]\int_0^{\infty}\frac{\rm{d}\Omega}{2\pi}|\chi[\Omega]|^2$, it follows that $\langle X_{\rm{sig}}^2\rangle\ll\langle X_{\rm{th}}^2\rangle$, and we can write $\Delta s(\tau)\simeq\Delta s_{\mathrm{th-th}}(\tau)$: Averaging $X^2$ along $\tau$ enables detecting a minimum excess of energy $\Delta\mathcal{E}=\frac{1}{2}m_{\rm{eff}}\Omega_{\rm{m}}^2\Delta s_{\mathrm{th-th}}(\tau)$, related to a minimum detectable force with a SD $S_{\rm{FF}}^{\mathrm{sig,min}}[\Omega_{\rm{m}},\tau]$ given by:

\begin{eqnarray}
S_{\rm{FF}}^{\mathrm{sig,min}}[\Omega_{\rm{m}},\tau]&=&\frac{\Delta s_{\rm{th-th}}(\tau)}{\int_0^{\infty}\frac{\rm{d}\Omega}{2\pi}|\chi[\Omega]|^2}\nonumber\\
&=&\sqrt{2}H(\Gamma_{\rm{M}},\tau)S_{\rm{FF}}^{\mathrm{th}}[\Omega_{\rm{M}}].
\label{eq:accuracy}
\end{eqnarray}

Finally, we define the equivalent force spectral resolution (EFSR, in $\mathrm{aN}/\sqrt{\rm{Hz}}$ units) as the square root of the minimum detectable force SD :
\begin{equation}
\delta F[\Omega_{\rm{m}},\tau]=\sqrt{S_{\rm{FF}}^{\mathrm{sig,min}}[\Omega_{\rm{m}},\tau]}.\label{eq:EFSR}
\end{equation}
Equations \ref{eq:accuracy} and \ref{eq:EFSR}  show that the time evolution of the EFSR is fully described by the function $H(\Gamma_{\rm{M}},\tau)$. For $\tau\ll1/\Gamma_{\rm{M}}$ the EFSR stays constant due to the mechanical memory of the transducer: averaging over a time smaller than the mechanical coherence time is completely inefficient, corroborating the advantage of using high frequency, high-$Q_{\rm{M}}$ oscillators. For $\tau\gg1/\Gamma_{\rm{M}}$, the EFSR varies as $(\Gamma_{\rm{M}}\tau)^{-\frac{1}{4}}$ confirming previous reports \cite{Rugar2004}. The discussion above shows that obtaining a given force resolution becomes faster as the damping rate of the oscillator increases. As dissipative feedback can provide an effective damping rate significantly larger than the intrinsic one, while keeping the thermal force unchanged, it enables substantial decrease of the averaging time.  In presence of dissipative feedback, the EFSR is 

\begin{eqnarray}
	\delta F[\Omega_{\rm{m}},\Gamma_{\rm{eff}},\tau]&=&\sqrt{H(\Gamma_{\rm{eff}},\tau)/H(\Gamma_{\rm{M}},\tau)}\times\delta F[\Omega_{\rm{m}},\tau] \nonumber\\
	& \simeq & \sqrt[4]{\Gamma_{\rm{M}}/\Gamma_{\rm{eff}}}\times\delta F[\Omega_{\rm{m}},\tau], 
\end{eqnarray}

where $\Gamma_{\rm{eff}}$ and $\delta F[\Omega_{\rm{m}},\tau]$ are given by Equations \ref{eq:effective-gamma} and \ref{eq:accuracy}. Consequently and importantly, feedback control reduces the time needed to reach a given force resolution by a factor of $\Gamma_{\rm{eff}}/\Gamma_{\rm{M}}$ which is of high practical relevance.

Figure 4(a) shows the trajectory of mechanical motion in its phase-space, without and in presence of dissipative feedback (blue and purple traces, respectively). Figure 4(b) shows the EFSR calculated as a function of $\Gamma_{\mathrm{{eff}}}$ and $\tau$ using
Equations \ref{eq:accuracy} and \ref{eq:EFSR}. Figure 4(c) shows the
corresponding experimental results. The contours correspond to plane regions
with constant resolutions described by $(\Gamma_{\mathrm{{eff}}}\tau\simeq
a_{i})_{a_{i} \in\mathbb{R}^{+}}$. The good agreement between theory and
experiment confirms that an increase of $\Gamma_{\mathrm{{eff}}}$ enables
decreasing $\tau$ by the same amount, while maintaining the same force
resolution: detecting a weak stationary incoherent force is 57 times faster in presence of a feedback rate $\Gamma_{\rm{eff}}/\Gamma=57$ (upper dashed line in Figure 4(c)) than without feedback (lower dashed line in Figure 4(c)). Figure 4(d) shows details of the EFSR time evolution corresponding to these two cases. The upper curve corresponds to
$\Gamma_{\mathrm{{eff}}}/\Gamma_{\mathrm{{M}}}=1$ (no feedback) and the lower
to $\Gamma_{\mathrm{{eff}}}/\Gamma_{\mathrm{{M}}}=57$. Both curves show a
scaling with $\tau^{1/4}$ in their asymptotic behavior (dashed gray lines), in excellent
agreement with theoretical expectations. A curvature is present in the upper curve at short
averaging times, which is absent in the lower one. We attribute this feature
to the mechanical memory time of the transducer $t_{\mathrm{{M}}}%
=1/\Gamma_{\mathrm{{M}}}\simeq26\,\mathrm{{ms}}$ as described by the
expression of $H(\Gamma_{\mathrm{{M}}},\tau)$ given above. Figure
4(e) shows the short term evolution of the force resolution in absence of
feedback (framed region in Fig. 4(d)) together with our theoretical model (Eqs. \ref{eq:accuracy} and \ref{eq:EFSR}, see
Methods) with no adjustable parameter being used. Excellent agreement is again found
between measurement and theory, thus validating our model for weak force
detection. Fig. 4(d) gives the improvement of the EFSR as a
function of $\Gamma_{\mathrm{{eff}}}/\Gamma_{\mathrm{{M}}}$, for a given
averaging time of $\tau=0.3\,\mathrm{{s}}$ as given along the vertical
dashed line in Fig. 4(b). The straight line corresponds to a power-law fit
with an exponent found to be $0.27$, in agreement with a value of $1/4$ expected
from the theoretical discussion.\\

\noindent \textbf{Feedback assisted incoherent force detection.}
We now explicitly show how feedback allows to enhance the system's force resolution by detecting a weak, incoherent radiation pressure force. Particularly, we show that feedback assisted force detection may even allow the detection of incoherent signals that would remain out of reach under standard operating conditions. We apply the force to be detected by sending a third laser tone into the tapered fiber which has a randomly modulated amplitude (see Fig.2(a) and Methods). It is important to emphasize that feedback enables an improved force resolution only for signals with coherence times being small compared to the effective mechanical decay time $2\pi/\Gamma_{\mathrm{eff}}$ (i.e. spectrally broad signals c.f. Fig.5(a)). By measuring the autocorrelation function of our signal, we determined its coherence time to be more than three orders of magnitude smaller than the intrinsic mechanical decay time $2\pi/\Gamma_{\mathrm{M}}$ (see Fig.5(b) and Methods), so that feedback gains up to a few hundreds can be safely employed. To detect the signal we use the "on/off" protocol suggested above: We first average the probe phase fluctuations while the signal is being applied and compare these to the averaged probe fluctuations in absence of the signal. One critical limit of this method results from sensitivity fluctuations giving an upper boundary for the possible averaging time. With our system we determined the sensitivity drifts to be on the order of $1\%$ at the minute scale: Signals with $S_{\mathrm{FF}}^{\mathrm{sig}}[\Omega\simeq\Omega_{\mathrm{M}}]\leq0.01\times S_{\mathrm{FF}}^{\mathrm{th}}[\Omega\simeq\Omega_{\mathrm{M}}]$ can therefore only be detected within less than one minute of averaging using the on/off protocol. Reducing the required acquisition time via feedback is therefore crucial, as it enables breaking the sensitivity limits induced by the stability constraints of the system.
We implemented the on/off protocol for various signal powers, feedback gains, and averaging times. Figure 5 shows signal detection for three different power levels, deduced from the signal's autocorrelation functions shown in Fig.5(c). Figs. 5(d,e,f) show the estimated energy excess induced by the presence of the signal $(s(\tau)-\mathcal{E}_{\mathrm{th}})/\mathcal{E}_{\mathrm{th}}$ (expressed in units of thermal energy $\mathcal{E}_{\mathrm{th}}$) as a function of averaging time $\tau$ and feedback gain. The color of the points is related to the averaging time, varying between $0\,\mathrm{s}$ and $35\,\mathrm{s}$ from blue to red. For each signal power the estimated energy excess converges towards the expected value (dashed blue lines, inferred from the level of the autocorrelation), with the final dispersion being in excellent agreement with our model (pink lines, see Methods). In particular, it is clearly confirmed that employing high feedback gains enables a significant decrease of the measurement imprecision for a given averaging time. Finally, we focus on the detection of the signal with the lowest power ($S_{\mathrm{FF}}^{\mathrm{sig}}[\Omega\simeq\Omega_{\mathrm{M}}]\simeq0.04\times S_{\mathrm{FF}}^{\mathrm{th}}[\Omega\simeq\Omega_{\mathrm{M}}]\simeq(15\,\mathrm{aN}/\sqrt{\mathrm{Hz}})^2$), with the results shown in Figs.5(f) and 5(g). The pink region in Fig.5(f) represents what we define as the Detection Zone (DZ) highlighting the points located within five standard deviations of the expected value of the measurement. The signal is considered to be detected in a given averaging time after which the centered normalized energy estimator (CNE) $(s(\tau)-\mathcal{E}_{\mathrm{th}})/\mathcal{E}_{\mathrm{th}}$ stays confined in the DZ. Fig.5(g) shows the detailed time evolution of the CNE given by the framed regions in Fig.5(f). In absence of feedback (Fig.5(g)(i)), $35\,\mathrm{s}$ of averaging is clearly insufficient to make a conclusion about the presence of the signal. Setting a moderate feedback gain $g=6.5$ (Fig.5(g)(ii)) reduces the measurement dispersion to $2$ standard deviations after $35\,\mathrm{s}$ of averaging, but is still insufficient to decide upon signal detection. Fig.5(g)(iii) shows that employing a relatively high gain $g=32$ enables detecting the signal, i.e. the applied incoherent force, within the same $35\,s$ of averaging duration. Note that reaching this resolution without the assistance of feedback would require averaging $32$ times longer amounting to approximately $20\,\mathrm{min}$. However, the sensitivity drifts in our system make averaging inefficient for this timescale. Thus, feedback is the condition \textit{sine qua non} in this experiment.\\

\noindent \textbf{Conclusion.}
In conclusion, we have realized an integrated hybrid force transducer system for sensing applications of ultrasmall signals. The integration of the system is thereby of crucial importance, as it allows a stable operation over an extended period of time. Otherwise, even the smallest changes in the system, particularly regarding the evanescent coupling between mechanical beam and the optical cavity, would prohibit the detection of small signals well below the thermal noise. In contrast to other high-$Q_{\rm{{M}%
}}$ and ultra-sensitive transducers \cite{Regal2008,LaHaye2004}, our system
operates at room temperature and profits from well-established read-out
techniques at optical frequencies. Moreover, our system also strongly benefits from the optomechanical interaction, as it enables to use multiple optical modes as channels for interaction (readout and actuation) with the nanomechanical beam. Concurrently, the integrated system could
be also implemented in a cryogenic environment in a straightforward manner
\cite{Riviere2011}.  The proposed and experimentally implemented
scheme to improve the force sensitivity for incoherent signals via
dissipative feedback has the potential to significantly reduce measurement
times for detection of weak signals. This scheme could be potentially applied for
magnetic resonance force microscopy \cite{Rugar2004} and other signals such as
transitions in nitrogen-vacancy (NV) centers
\cite{Balasubramanian2008,Maze2008}. The system can be readily functionalised, for example via the use of nanomanipulators \cite{vanderSar2009} or using a scanning atomic force microscope \cite{Schroder2009,Schell2011}. Finally, the integrated system is a
promising candidate for experiments in optomechanics \cite{Kippenberg2008} due to its high $Q_{\mathrm{{M}}}$, potential of
cryogenic implementation and applicability to feedback cooling methods that do
not require the resolved sideband regime for ground state cooling
\cite{Courty2001}. The lowest occupancy achieved with our system at room
temperature is $\bar{n}\approx5000$ ($T_{\mathrm{eff}}=700\;\mathrm{mK}$)
giving a promising outlook for operation at cryogenic temperatures.

\vspace{0.2in}

\noindent \textbf{Methods}

{\footnotesize \noindent \textbf{Fabrication} The samples are fabricated on a silicon
wafer with thermally grown 2 $\mu\mathrm{m}$ silica and 100 nm LPCVD deposited
high-stress stoichometric Si$_{3}$N$_{4}$ thin films on top. First, the
nanobeam is created using electron-beam lithography followed by a timed
C$_{4}$F$_{8}$-SF$_{6}$ chemistry reactive ion etching. In a second
lithography step, the polymer mask for the disk resonator is defined in such
a way that the disk partially overlaps the nanobeam. The disk resonator was
created via a wet etch in a buffered hydrofluoric bath. Due to the isotropic nature of
this etch, material below the polymer mask is partially etched resulting in an
angled sidewall of the disk. Using this effect and ensuring the correct
alignment of the disk mask with respect to the nanobeam, the release of the
nanobeam is achieved with the structure being positioned above the wedge of
the disk resonator. Release of the combined system is accomplished by
anisotropic silicon etching in a potassium hydroxide bath.}

{\footnotesize \vspace{0.1in} }

{\footnotesize \noindent \textbf{Actuation of the nanomechanical oscillator} A
pump-probe scheme is used to actuate and read out the motion of the
nanomechanical oscillator. Both the pump and the probe light fields are
provided by external cavity diode lasers (ECDL). The probe beam is phase
modulated using an electro-optical modulator (EOM) and the pump is amplitude
modulated via a LiNbO$_{3}$ Mach-Zehnder modulator (AM). Both beams are
combined with a fiber-based beam splitter (C1) and coupled to two different optical
modes using a tapered fiber. Both optical modes belong to the same WGM mode
family and are separated by one free spectral range. The light coupled back
from the resonator to the tapered fiber is split by a second fiber based beam
splitter (C2) and tunable optical filters (TOF1 and TOF2) are used to separate the spectral
signatures of the pump and the probe beam. Light is detected via fast photo
receivers (PD1 and PD2) and analyzed with an oscilloscope as well as an electrical
spectrum analyzer (ESA) or network analyzer (NA). The frequency of both lasers
is locked on the side of the fringe of the respective optical modes by
providing an error signal to the piezo controller of the laser. In case of the
ringdown measurement, the pump is amplitude modulated at the mechanical
resonance frequency by providing to the AM the corresponding radio frequency signal from
the NA set to zero span mode. The detected signal is analyzed by the
NA. After interrupting the modulation a temporal decrease of the oscillation amplitude can be
observed from which the mechanical $Q$ is obtained. In case of the response
measurement, the signal provided to the AM from the NA is swept around the
mechanical resonance frequency and the response of the system is
demodulated and recorded by the NA. The sweeps are always conducted from smaller
to larger frequencies. }

{\footnotesize \vspace{0.1in} }

{\footnotesize \noindent \textbf{Implementation of the feedback mechanism} The
experimental setup is in principle the same pump-probe scheme as described
above, however with the difference that the signal fed to the AM is derived
from quadrature detection. Quadrature detection is accomplished by
splitting the signal from the PD detecting the light of the probe beam and
demodulating it. The demodulation is achieved by mixing (low-noise analog mixer {\sc Minicircuits} ZP-3-S+) both arms of the
splitted signal with a radio frequency source (at the mechanical resonance
frequency $\Omega_{M}/2\pi$) from a signal generator ({\sc Tektronix} AFG3102) with
the phase of the source delivered to both arms being shifted by 90$^{\circ}$.
The corresponding signals are filtered with a low-pass (3 dB cutoff at 300 kHz) and
give the time evolution of both quadratures. To obtain the error signal for
the feedback, both low-pass filtered signals are modulated at the mechanical
resonance frequency by a second signal generator with the phase being again
shifted by 90$^{\circ}$ for both arms. The phase between both signal
generators is locked to a 10 MHz voltage controlled oscillator. Adjusting
the phase difference corresponds to adjusting the phase $\Phi$ of the feedback
signal. Both signal arms are combined after the modulation and after a low
noise amplifier (Miteq 1447) fed to the AM. }

{\footnotesize \vspace{0.1in} }

{\footnotesize \noindent \textbf{Enhancement of force resolution using feedback} To
determine the measurement accuracy (Eq. \ref{eq:dispersions1}) corresponding
to a given feedback gain, we first extract the quadratures of motion by
sending the detected signal to a spectrum analyzer ({\sc Agilent}$\,$MXA$\,$9020A),
operating in $I/Q$ mode with a sampling rate of $1/\tau_{\mathrm{{s}}}%
\simeq7\,\mathrm{{kHz}}$. We record independent realizations of the thermal
noise $(X_{1}(t),X_{2}(t))_{i,j}$ along $N$ different acquisition times
$\tau_{j}$ ranging from $27\,\mathrm{{ms}}$ to $3.77\,\mathrm{{s}}$. Each of
the resulting sequences is used in order to determine the corresponding
averaged energy $s_{i,j}(\tau_{j})=1/\tau_{j}\int_{0}^{\tau_{j}}%
\mathrm{d}t\left( X_{1,(i,j)}^{2}(t)+X_{2,(i,j)}^{2}(t)\right) $. The
measurement resolution after averaging time $\tau_{j}$ is then determined as
$\Delta s(\tau_{j})=\sqrt{\mathrm{Var}\left[ \left( s_{i,j}(\tau_{j})\right)
_{i}\right] }$. We define the equivalent force resolution (represented in
Fig.4) as the equivalent motion resolution normalized to the area of the
mechanical amplitude response $\delta F(\tau_{j})=\sqrt{\Delta s(\tau
_{j})/\int_{0}^{\infty}\frac{\mathrm{{d}\Omega}}{2\pi}|\chi[\Omega]|^{2}}$,
the latter being determined by measuring the expected value of the energy
$\langle s(\tau_{j})\rangle\simeq1/N\sum_{j=0}^{N}s(\tau_{j})$, using that
$\langle s(\tau)\rangle=S_{\mathrm{FF}}^{\mathrm{th}}[\Omega_{\mathrm{M}}%
]\int_{0}^{\infty}\frac{\mathrm{{d}\Omega}}{2\pi}|\chi[\Omega]|^{2}$. }

{\footnotesize \vspace{0.1in} }

{\footnotesize \noindent \textbf{Feedback assisted force detection} To obtain the incoherent radiation pressure force, the output laser field of a third ECDL is sent into an AM fed with a Gaussian white noise generated by a function generator ({\sc Agilent}  $\mathrm{33250A}$). To avoid saturating the AM, the generator output is filtered using a selective $3\,\mathrm{MHz}$ bandwidth low-pass filter. The optical frequency of this ECDL is tuned in order to match a third optical resonance of the WGM cavity, distinct from the one used to probe (ECDL1) and to actuate (ECDL2) the feedback response. This third "noisy" tone is sent into the tapered fiber using a fiber based beam splitter (C3 on Fig.2(a)) and is later separated from the other two tones using the combination of another beam splitter (C4) and of an optical filter (TOF3). A photodetector (PD3) enables measuring the transmitted intensity fluctuations. To determine the autocorrelations functions presented in Fig. 5(b), we proceeded in the following manner: The time evolution of the signal's quadratures is first recorded for a duration $T_{\mathrm{s}}\simeq35\,\mathrm{s}$ at a sampling rate $\mathrm{d}t\simeq0.2\,\mathrm{ms}$ using an ESA operating in I/Q mode. The autocorrelation being quadrature-invariant for Gaussian signals, only one quadrature $X$ is conserved for its computation, which we restricted to delay times $\delta\tau$ smaller than $\delta\tau_{\mathrm{max}}=30\,\mathrm{ms}$. For any $\delta\tau\in[-\delta\tau_{\mathrm{max}},\delta\tau_{\mathrm{max}}]\cap \mathrm{d}t\times\mathbf{N}$ the autocorrelation value $\langle X(0)X(\delta\tau)\rangle$ is obtained by evaluating the discrete time integral $\langle X(0)X(\delta\tau)\rangle\simeq\frac{1}{T_{\mathrm{s}}-2\delta\tau_{\mathrm{max}}}\int_{\delta\tau_{\mathrm{max}}}^{T_{\mathrm{s}}-\delta\tau_{\mathrm{max}}}\mathrm{d}t X(t)X(t+\delta\tau)$, the latter equality applying by virtue of the ergodic theorem. As both signals being measured (the intensity noise and the thermal displacement of the nanomechanical oscillator) are resulting from white noises, their autocorrelation function is expected to be the impulse response of the measurement apparatus used to detect them (\{nanomechanical oscillator+WGM resonator+photodetector+ESA\} and \{WGM resonator+photodetector+ESA\} for the mechanical motion and the intensity noise, respectively). For the probe signal, this impulse response is mostly due to the mechanical response and can be approximated by a decaying exponential, since the sampling rate is $\mathrm{d}t\ll2\pi/\Gamma_{\rm{M}}$. For the intensity noise, this impulse response prominently involves the $5\,\mathrm{kHz}$  wide frequency gate set by the I/Q mode parameters of the ESA, such that the autocorrelation linewidth in Figs. 5(b,c) is not related to the coherence time of the intensity noise but simply to this frequency gate. The pink dispersion lines in Fig. 5(f) are determined using that $\Delta\mathcal{E}(\tau)=\frac{1}{2}m_{\rm{eff}}\Omega_{\rm{m}}^2\Delta s_{\mathrm{th-th}}(\tau)$ since the low SNR hypothesis applies. This is, however, not the case in Figs. 5(d,e), where the SNR are $\langle X_{\mathrm{sig}}^2\rangle/\langle X_{\mathrm{th}}^2\rangle\simeq5$ and $\langle X_{\mathrm{sig}}^2\rangle/\langle X_{\mathrm{th}}^2\rangle\simeq0.5$ respectively. The corresponding dispersions are therefore evaluated using an extended model taking into account the terms $\Delta s_{\mathrm{sig-th}}(\tau)$ and $\Delta s_{\mathrm{sig-sig}}(\tau)$ (Eqs. \ref{eq:dispersions1} and \ref{eq:dispersions2}).}

\textbf{Acknowledgements.}  The fabrication was carried out in the Center of MicroNanotechnology (CMi) at EPFL. We acknowledge financial support by the NCCR Quantum Photonics and the DARPA Orchid program.

\newpage
\widetext
\newpage

\begin{figure}[ptb]
\includegraphics[scale=1.95]{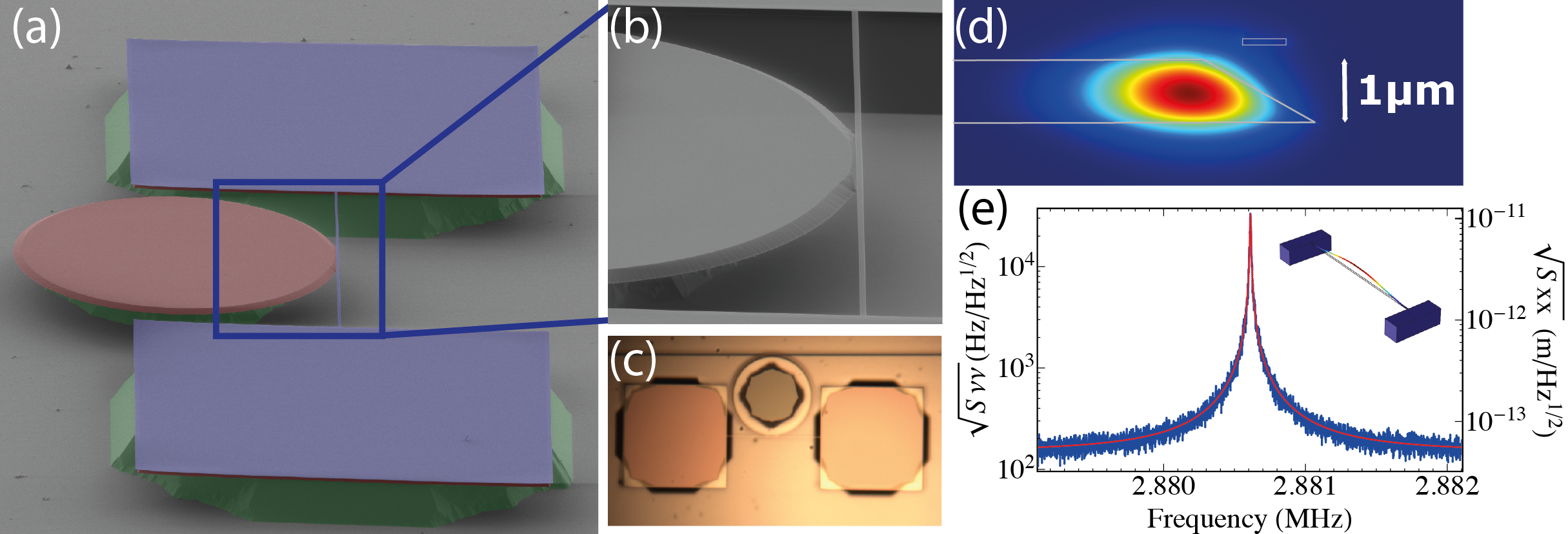}
\caption{\textbf{Hybrid nanomechanical transducer system on a  chip.} (a) Scanning electron image (false colors) of the hybrid on-chip system consisting of a doubly-clamped, high-stress Si$_{3}$N$_{4}$ nanomechanical beam (blue) near-field coupled to a silica disk resonator (red). Both, the resonator as well as the pads holding the beam rest on silicon pedestals (green). (b) Magnified image of the proximity area. (c) Optical micrograph of a tapered fiber coupled to the hybrid system. (d) Finite-element simulation of the electromagnetic field distribution of the fundamental optical mode confined in the disk resonator. The nanomechanical beam is coupled to the evanescent near-field of the optical mode. (e) Room-temperature Brownian noise of a nanomechanical beam with a fundamental resonance frequency of 2.88 MHz and $m_{\rm{eff}}=9\;\mathrm{pg}$. Fitting the mode gives a mechanical quality factor of $Q_{\rm{M}}=4.3\times 10^{5}$. Inset: Finite-element simulation of the beam's fundamental out-of-plane mode.}%
\label{Fig1}%
\end{figure}

\clearpage

\begin{figure}[ptb]
\includegraphics[scale=1.35]{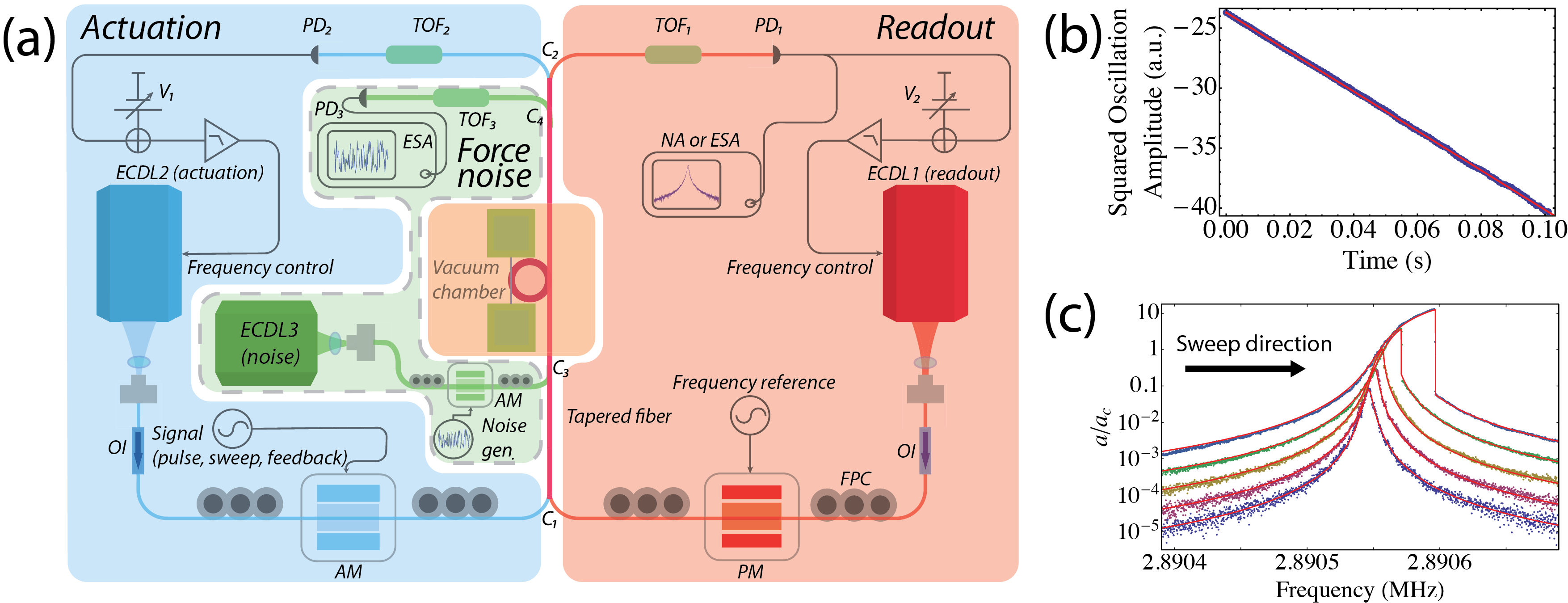}
\caption{\textbf{Actuation of the nanomechanical oscillator by radiation pressure.} (a) Scheme of the actuation-readout setup used for the actuation and feedback experiments. By amplitude modulating a third laser (green shaded area) an incoherent force can be applied on the transducer for direct force detection experiments. Details of the setup are given in Methods. (b) Ringdown measurement of the fundamental mode fitted with an exponential decay function (red line) confirming the high $Q_{\rm{M}}$. (c) Response measurements for an increasing actuation of the nanomechanical motion. The actuation power is increased by 5 dB for the curves from bottom to top. The direction of the sweep is from lower to higher frequencies. With increasing actuation the response of the oscillator is increased and above a certain threshold a mechanical bistability sets in. The middle curve (dark yellow) was taken at the threshold and is used to calibrate the actuation in terms of the critical displacement $a_{c}$. $a_{c}$ is found to be 1.6 nm for our system.}%
\label{Fig2}%
\end{figure}

\clearpage

\begin{figure}[ptb]
\includegraphics[scale=1]{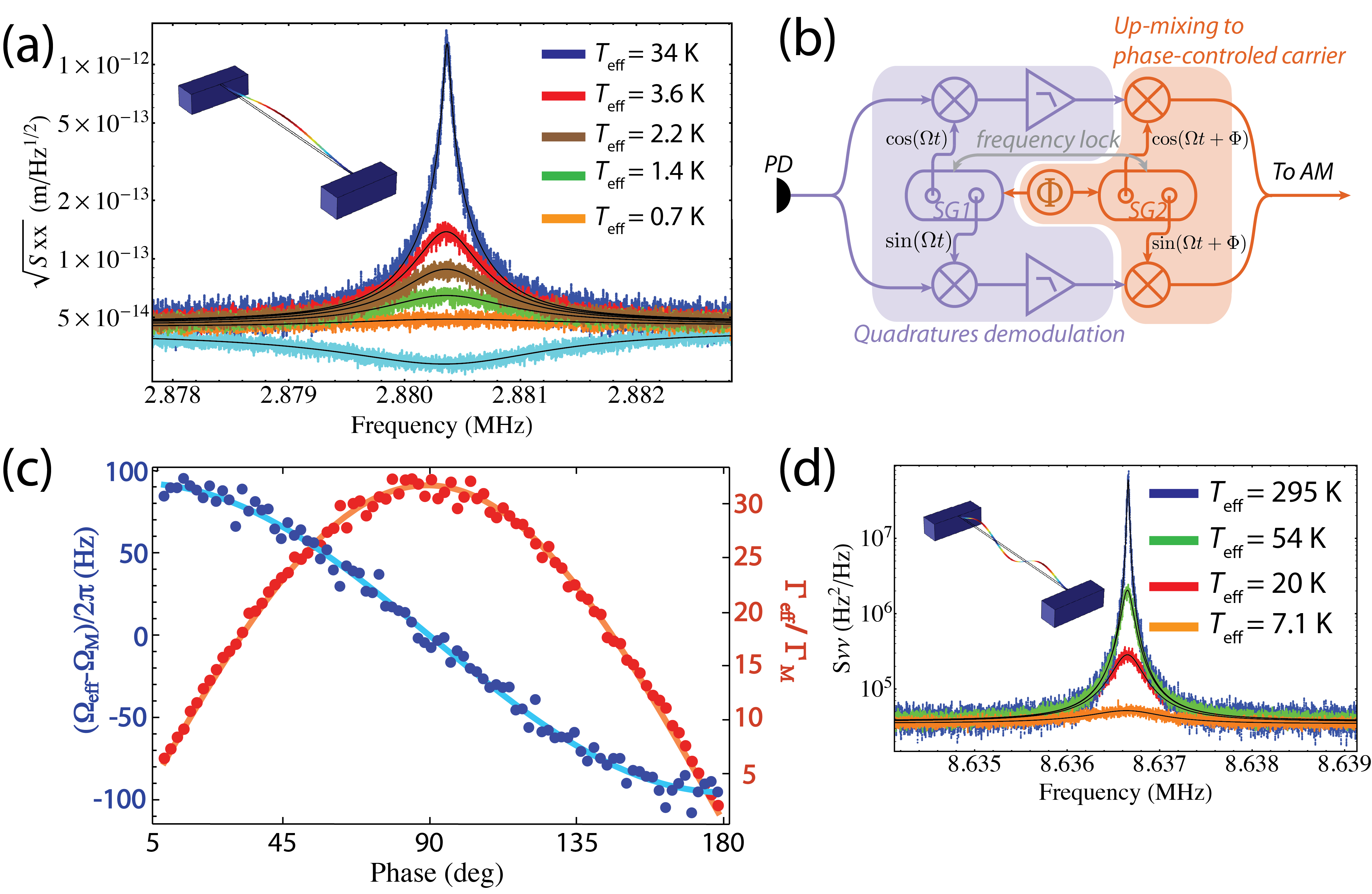}
\caption{\textbf{Feedback control of the nanomechanical transducer.} (a) Feedback cooling of the fundamental out-of-plane mode. The light blue curve corresponds to a squashing of the background present. While it can still be fitted our model, we do not use it to
determine $T_{\mathrm{eff}}$. Inset: Finite-element simulation of the beam's fundamental out-of-plane mode. (b) Feedback scheme based on demodulation and modulation of the incoming probe signal (Details see text and Methods). (c) Feedback cooling of the fundamental in-plane mode. Inset: Finite-element simulation of the beam's fundamental in-plane mode. (d) Feedback cooling of the third order harmonic of the out-of-plane mode family. Inset: Finite-element simulation of the beam's third order out-of-plane mode. (e) Frequency detuning $\left(\Omega_{\mathrm{eff}}-\Omega_{\rm{M}}\right)/2\pi$ of the fundamental out-of-plane mode (left axis) and ratio of the effective linewidth to the intrinsic linewidth (right axis) versus the phase difference $\Phi$ between the demodulation and modulation signals. }%
\label{Fig3}%
\end{figure}

\clearpage

\begin{figure}[ptb]
\includegraphics[scale=0.85]{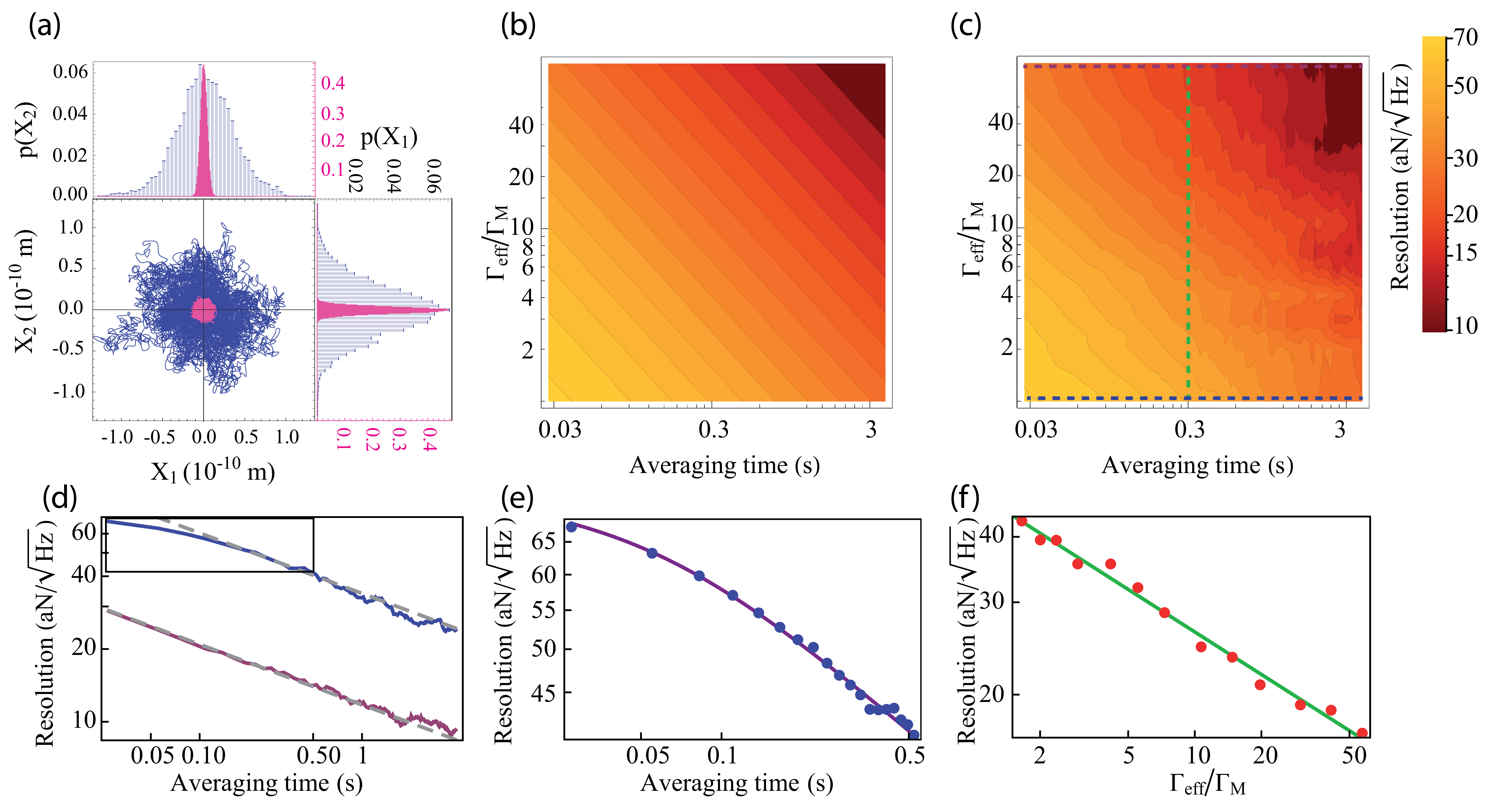}
\caption{\textbf{Force resolution enhancement via feedback control.}  (a) Evolution of mechanical motion in phase-space without  and in presence of feedback (blue trace and purple traces, respectively). The histograms correspond to the statistical distributions of quadratures $X_1$ and $X_2$. (b) Calculated force resolution as a function of the effective damping $\Gamma_{\rm{eff}}$ and average time $\tau$ and (c) corresponding experimental result. (d) Evolution of the force resolution as a function of $\tau$ without feedback (upper curve, corresponding to the blue dashed line in Fig. 5(c)) and with $\Gamma_{\rm{eff}}/\Gamma_{\rm{M}}=57$ (lower curve, corresponding to the purple dashed line in Fig. 5(c)). Both curves show very good agreement with an asymptotic behavior scaling with $\tau^{1/4}$ (dashed gray lines). (e) Short term evolution corresponding to the framed region in (d) together with the theoretical model. (f) Evolution of force resolution as a function of $\Gamma_{\rm{eff}}/\Gamma_{\rm{M}}$ for a fixed averaging time of $\tau=0.3\,\rm{s}$, corresponding to the evolution along the vertical dashed line in Fig. 5(b).}%
\label{Fig4}%
\end{figure}

\begin{figure}[ptb]
\includegraphics[scale=1]{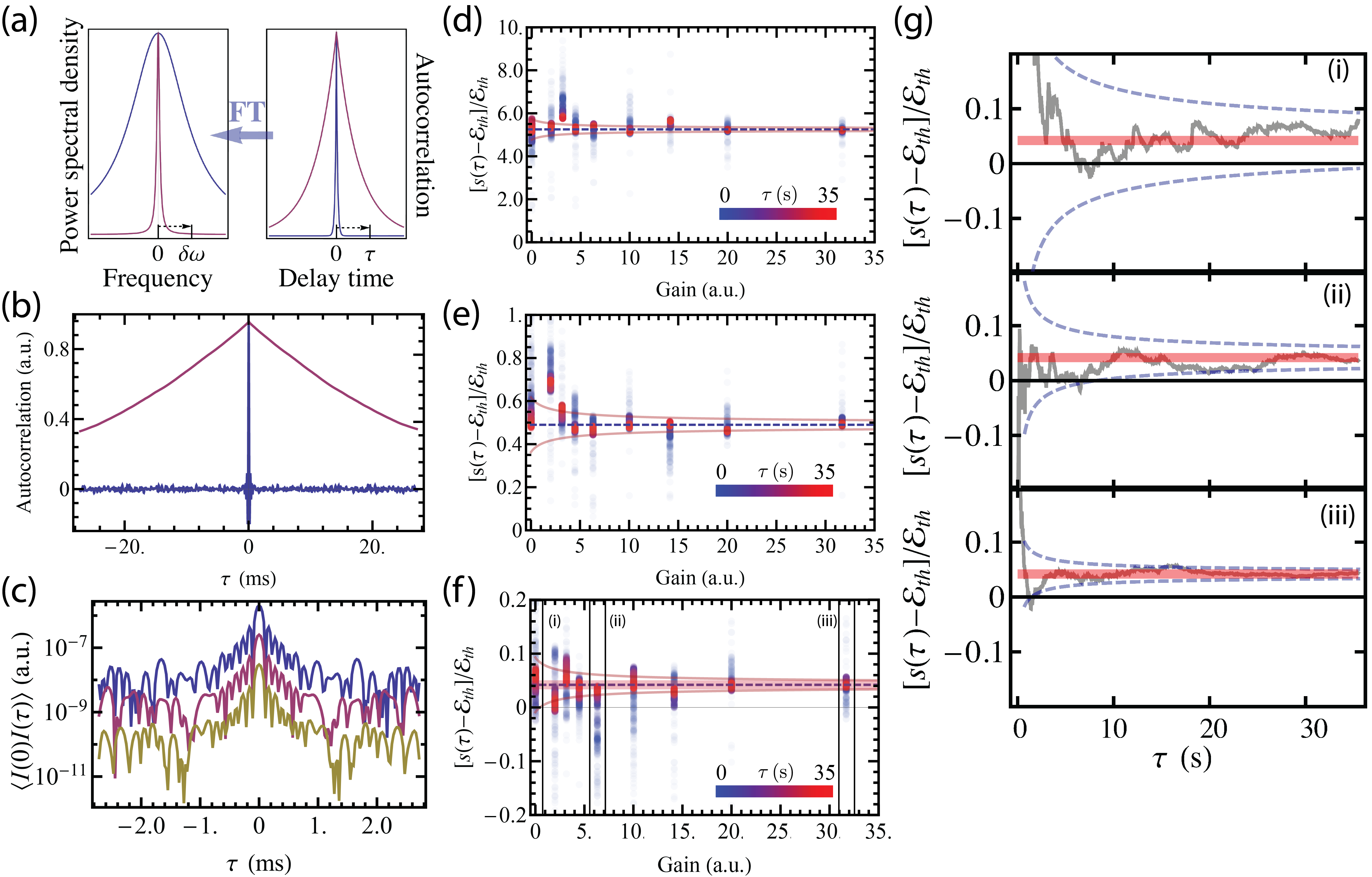}
\caption{\textbf{Feedback-assisted force detection.}  (a) Scheme illustrating the validity limit for the feedback-assisted weak random force detection. Left: The quadrature spectrum of the driving signal (in blue) has to be broad when compared to the mechanical spectrum (in purple). Right: Equivalently, the autocorrelation function of the signal (linked to its spectrum via a Fourier Transform (FT) operation) has to be narrow as compared to the autocorrelation function of the mechanical motion. (b) Experimentally measured autocorrelation functions associated with the signal intensity noise (in blue) and with the thermally driven mechanical motion (in purple). (c) From top to bottom: Intensity noise autocorrelation function for decreasing noise powers, associated with the signals detected in (d), (e) and (f) respectively. (d-f) Signal contribution to the mechanical energy (expressed in units of thermal noise), measured along various averaging times and feedback gains. Each point corresponds to the signal contribution obtained in presence of a feedback gain given by its abscise value and after a given averaging time related to its color (single shot measurement for the bluer poins, $35\,\mathrm{s}$ of averaging for the redder ones). The blue dashed lines correspond to the expected values of the signal contributions to the mechanical energy ($5.25\times\mathcal{E}_{\mathrm{th}}$, $0.5\times\mathcal{E}_{\mathrm{th}}$ and $0.04\times\mathcal{E}_{\mathrm{th}}$, respectively), deduced from the levels of the signal autocorrelation functions shown in Fig. 6(c). The pink lines correspond to the gain evolution of the energy dispersion expected after $35\,\mathrm{s}$ of averaging (see Methods). The pink region in (f) gives the signal detection zone, which we define as corresponding to a measurement accuracy of $5$ standard deviations or better. (g) Figs. 6(g)(i), 6(g)(ii) and 6(g)(iii) represent the detailed evolution of the signal contribution with averaging time for feedback gains of $0$, $6.5$ and $32$ respectively for the lowest signal applied, $\mathcal{E}_{\mathrm{sig}}\simeq0.04\,\mathcal{E}_{\mathrm{th}}$. The blue dashed line corresponds to the expected measurement dispersion (see Methods). The pink region corresponds to the detection zone.}%
\label{Fig5}%
\end{figure}

\end{document}